\documentclass[10pt,letterpaper]{article}
\usepackage{opex3}
\usepackage{cite}
\usepackage{amsmath,amssymb,graphicx}
\usepackage{xfrac}
\usepackage{hyperref}
\usepackage{cleveref}
\usepackage[none]{hyphenat}
\begin{document}

\title{Deterministic single soliton generation and compression in microring resonators avoiding the chaotic region}

\author{Jose A. Jaramillo-Villegas,$^{1,3,*}$ Xiaoxiao Xue,$^1$ Pei-Hsun Wang,$^1$ Daniel E. Leaird,$^1$ and Andrew M. Weiner$^{1,2}$}

\address{$^1$School of Electrical and Computer Engineering, Purdue University, West Lafayette, Indiana 47907, USA \\
$^2$Birck Nanotechnology Center, Purdue University, West Lafayette, Indiana 47907, USA \\
$^3$Facultad de Ingenier\'{i}as, Universidad Tecnol\'{o}gica de Pereira, Pereira, Risaralda 660003, Colombia}

\email{$^*$jjv@purdue.edu} 

\begin{abstract}
A path within the parameter space of detuning and pump power is demonstrated in order to obtain a single cavity soliton (CS) with certainty in SiN microring resonators in the anomalous dispersion regime. Once the single CS state is reached, it is possible to continue a path to compress it, broadening the corresponding single free spectral range (FSR) Kerr frequency comb. The first step to achieve this goal is to identify the stable regions in the parameter space via numerical simulations of the Lugiato-Lefever equation (LLE). Later, using this identification, we define a path from the stable modulation instability (SMI) region to the stable cavity solitons (SCS) region avoiding the chaotic and unstable regions.
\end{abstract}

\ocis{(230.5750) Resonators; (190.5530) Pulse propagation and temporal solitons; (190.4380) Nonlinear optics, four-wave mixing.} 

\bibliographystyle{osajnl}
\bibliography{CAT20150328}

\section{Introduction}
An optical frequency comb is a light source with a number of highly resolved and nearly equidistant spectral lines. Since its introduction, multiple important applications have been demonstrated in areas such as communications, metrology, spectroscopy, astronomy and optical clocks. Optical frequency combs can be generated using mode-locked lasers or electro-optic modulation of continuous-wave light. Since 2007, multiple experiments have reported optical frequency comb generation by means of Kerr nonlinearity (wave mixing process) in microresonators, which offer potential for highly compact and portable solutions. Such combs are termed Kerr optical frequency combs or simply Kerr combs\cite{kippenberg2011microresonator, papp2011spectral, del2007optical, grudinin2009generation, levy2009cmos, razzari2009cmos, ferdous2011spectral, del2011octave, foster2011silicon, okawachi2011octave, wang2012observation}.

The understanding of underlying processes and dynamics in Kerr comb generation is critical to move this technology further to industry and applications. Chembo and Yu in \cite{chembo2010modal} introduced one of the first simulation approaches using modal expansion. More recently, the Lugiato-Lefever equation (LLE) \cite{lugiato1987spatial} has been widely adopted \cite{coen2013modeling, chembo2013spatiotemporal, coen2013universal, hansson2013dynamics, godey2013stability, balakireva2013stability, lamont2013route, coillet2013azimuthal, erkintalo2014coherence, parra2014dynamics, coillet2014robustness}. Furthermore, Chembo and Menyuk in \cite{chembo2013spatiotemporal} demonstrated the equivalence between the LLE and mode coupling equations models and Hansson et al. in \cite{hansson2014numerical} showed that both models can be numerically solved in similar computational times.

The generalized mean-field LLE equation is:
\begin{equation}
        t_R\frac{\partial E(t,\tau)}{\partial t} = \left[ -\alpha-i\delta_0+iL\sum_{k\ge2}\frac{\beta_k}{k!}\left(i \frac{\partial}{\partial\tau}\right)^k +i\gamma L|E|^2\right]E+\sqrt{\theta}E_{\textrm{in}} 
\label{eq:LLE}
\end{equation}
which is the nonlinear Schr\"{o}dinger equation (NLSE) with damping, detuning and external pumping, which accurately describes Kerr comb generation. In this equation $E(t,\tau)$ is the complex envelope of the total intracavity field and hereafter simply the field, $t$ is the so called slow time variable, $\tau$ is the fast time variable, $t_R$ is the round trip time, $\alpha$ is half of the total loss per round trip which includes the internal linear absorption and coupling loss, $\delta_0$ is the phase detuning, $L$ is the cavity length, $\beta_k$ is the $k$-order dispersion coefficient, $\gamma$ is the Kerr coefficient, $\theta$ is the coupling coefficient between the waveguide and the microresonator and $E_{\textrm{\fontsize{6}{1}\selectfont{in}}}$ is the pump field (normalized such that $P_{\textrm{\fontsize{6}{1}\selectfont{in}}}=|E_{\textrm{\fontsize{6}{1}\selectfont{in}}}|^2$). To facilitate our analyses in a more general framework we will additionally use the normalized detuning and pump field according to Eq. 2 and 3.
\begin{equation}
        \Delta=\frac{\delta_0}{\alpha} 
\label{eq:detuning}
\end{equation}

\begin{equation}
        S=E_{\textrm{\fontsize{6}{1}\selectfont{in}}} \sqrt{\frac{\gamma L\theta}{\alpha^3}}
\label{eq:pump}
\end{equation}
Recently, some authors explored and identified regions corresponding to different types of operation in the detuning and pump power $(\Delta,|S|^2)$ parameter space in the anomalous dispersion regime \cite{coen2013universal,balakireva2013stability,parra2014dynamics}. The characterized regions are: stable modulation instability (SMI), unstable modulation instability (UMI), stable cavity solitons (SCS), unstable cavity solitons (UCS) and continuous wave (CW). Additionally, Erkintalo and Coen in \cite{erkintalo2014coherence} explored the first-order coherence properties of each region.

Single CS generation is a desired goal in much research because it yields a high coherence, single free spectral range (FSR) Kerr comb. It has been suggested that this style of comb can be used as information carriers in optical communications and optical memories. Although the first experimental observation of a single CS in microresonators was reported in \cite{herr2014temporal}, the difficulty to control the number of CSs in the SCS region was also emphasized. In particular, in the usual approach in which detuning is swept to a predetermined final value with input power held constant, the number of CSs generated is probabilistic. In the current work, we show a method not only to obtain a single CS with a high degree of certainty avoiding the chaotic and unstable regions such as UMI and UCS, but also to compress it, moving the system through the SCS to high power values.

\section{Deterministic Single CS Generation}

The LLE is simulated numerically using the split step Fourier method (SSFM). The simulation parameters used in this work are: $t_R=1/226$ GHz, $\beta_2=-4.7\times10^{-26}$ s$^{2}$m$^{-1}$, $\alpha=0.00161$, $\gamma=1.09$ W$^{-1}$m$^{-1}$, $L=2\pi\times100$ $\mu$m and $\theta=0.00064$ \cite{wang2014coherent}. These parameters correspond to a Si$_3$N$_4$ microring resonator of 100 $\mu$m radius with anomalous dispersion, loaded quality factor $Q$ of $1.67\times10^6$ and photon lifetime $t_{\textrm{\fontsize{6}{1}\selectfont{ph}}} = t_R/(2\alpha) = 1.37$ ns. Addionally, the simulations are done initializing the intracavity field in the frequency domain $E(\omega)$ to a circularly symmetric complex Gaussian noise field with a standard deviation $\sigma_{\textrm{\fontsize{6}{1}\selectfont{noise}}} = 10^{-9}$ [W$^{1/2}$], which with normalization $P=|E|^2$ is equivalent to a mean power of -150 dBm per cavity mode and uniform random phase.

To show the probabilistic nature of the number of CS after a detuning swept at constant pump power we perform simulations according to the insets of Fig. 1(a). These insets depict how the detuning is swept linearly from $\Delta = 0$ to a final value of $\Delta_{\textrm{\fontsize{6}{1}\selectfont{f}}}$ over a time interval of 0.3 $\mu$s ($\approx$218 $t_{\textrm{\fontsize{6}{1}\selectfont{ph}}}$) at a constant $|S|^2$ = 18.9 ($P_{\textrm{\fontsize{6}{1}\selectfont{in}}}$ = 180 mW). Fig. 1(a) shows the number of temporal peaks (in the SCS region this corresponds to number of CSs) versus final value of detuning $\Delta_{\textrm{\fontsize{6}{1}\selectfont{f}}}$. The number of CSs obtained depends very sensitively on $\Delta_{\textrm{\fontsize{6}{1}\selectfont{f}}}$ even though the noise initialization was the same for all the simulations of Fig. 1(a). Additionally, the number of CSs is different when we repeat the simulation with the same detuning and pump parameters but different realizations of the initial noise, as shown in the histogram of Fig. 1(b). The behavior of the total intracavity energy $U_{\textrm{\fontsize{6}{1}\selectfont{Intra}}}$, which is proportional to the integral of the intensity over a round trip time, give us an idea of the Kerr comb stability. Fig. 1(c) shows total intracavity energy during a detuning swept at constant input power. The noisy behavior in the yellow segment gives us evidence that the system is in the UMI region. Then, the oscillatory behavior in the blue segment is a clear characteristic of UCS region. Finally, the total intracavity energy becomes stable when the microresonator approaches the SCS region. Our hypothesis in this work is that the uncertainty in the number of CSs in the SCS region results from chaotic field evolution while the system is crossing the UMI and UCS regions.

\begin{figure}[htpb]
\centerline{\includegraphics[width=31.5pc]{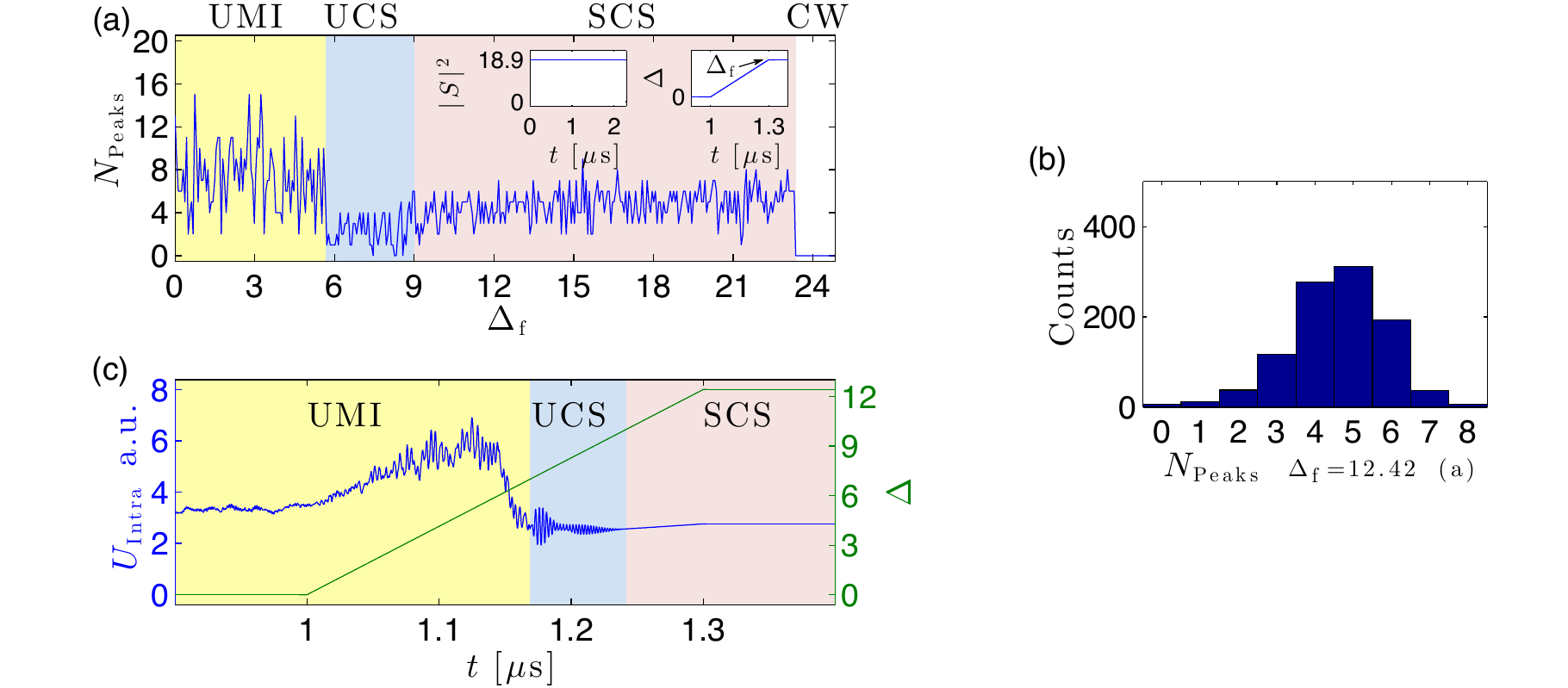}}
\caption{(a) The number of peaks as a function of final value of detuning $\Delta_{\textrm{\fontsize{6}{1}\selectfont{f}}}$ with $\Delta$ swept as shown in the insets. The pump power $|S|^2$ is set to a constant value of 18.9 ($P_{\textrm{\fontsize{6}{1}\selectfont{in}}}$ =180mW). These simulations were done using the same realization of initial noise. (b) Histogram of number of peaks for 1000 simulations with different realizations of initial noise, using the same detuning and pump parameters in the simulation of (a) with $\Delta_{\textrm{\fontsize{6}{1}\selectfont{f}}} = 12.42$. (c) Total intracavity energy (blue) and detuning (green) as a function of time for the simulation of (a) with $\Delta_{\textrm{\fontsize{6}{1}\selectfont{f}}} = 12.42$.}
\label{fig:Fig1}
\end{figure}

Although some of the boundaries between the regions can be solved analytically (e.g. Hamiltonian-Hopf limit between CW and SMI regions \cite{parra2014dynamics}), there is still no theoretical solution for the boundary between the SMI and UMI regions. Therefore, we characterize the $(\Delta,|S|^2)$ parameter space via numerical simulations to gain knowledge about these boundaries. We initialize the microresonator pump at a detuning $\Delta = 0$ and power $|S|^2 = 6.3$ (60 mW) with well-defined behavior in the SMI region. This point corresponds to the location of the red triangle in Fig. 2(a) where a Turing rolls pattern \cite{coillet2013azimuthal,coillet2014robustness} of 13 equally spaced peaks is formed. The intensity and power spectrum obtained at this point are shown in Fig. 2(c) (left) and (center), respectively. This pattern corresponds to a Kerr comb with frequency spacing between lines equal to 13 times the microresonator FSR and possesses low intensity noise and high stability, as shown by its steady state intracavity energy in Fig 2(c) (right). According to Erkintalo and Coen in \cite{erkintalo2014coherence}, the UMI region exhibits high first-order coherence. Coillet and Chembo in \cite{coillet2014robustness} demonstrated that Turing rolls patterns achieve phase locked states independently of the initial conditions and can be easily generated by decreasing the optical frequency of the pump laser towards the resonance, as presented in their experimental examples. These reasons make Turing rolls pattern a good choice for our initial point. After 200 ns, we jump in a single step to a point in $(\Delta,|S|^2)$ parameter space and keep the system there for 1.8 $\mu$s ($\approx$1310 $t_{\textrm{\fontsize{6}{1}\selectfont{ph}}}$). Simulations were repeated using the same realization of initial noise for different final values of detuning $\Delta$ and pump power $|S|^2$ up to 6.21 and 10.5 (100 mW), respectively. Fig. 2(a) shows the number of time domain peaks for each final point. Each point $(\Delta,|S|^2)$ is classified based on intensity, spectrum and total intracavity energy stability at the end of the simulation in order to determine the boundaries of the different operating regions. The characteristics of each region are shown in Fig. 2 (c) to (f) and the result of this classification is shown in Fig. 2(b). Notably, $U_{\textrm{\fontsize{6}{1}\selectfont{Intra}}}$ in the UMI and UCS regions is characterized by noisy and oscillatory behavior, respectively, while in the SMI and SCS regions it is constant as mentioned before.

\begin{figure}[htpb]
\centerline{\includegraphics[width=31.5pc]{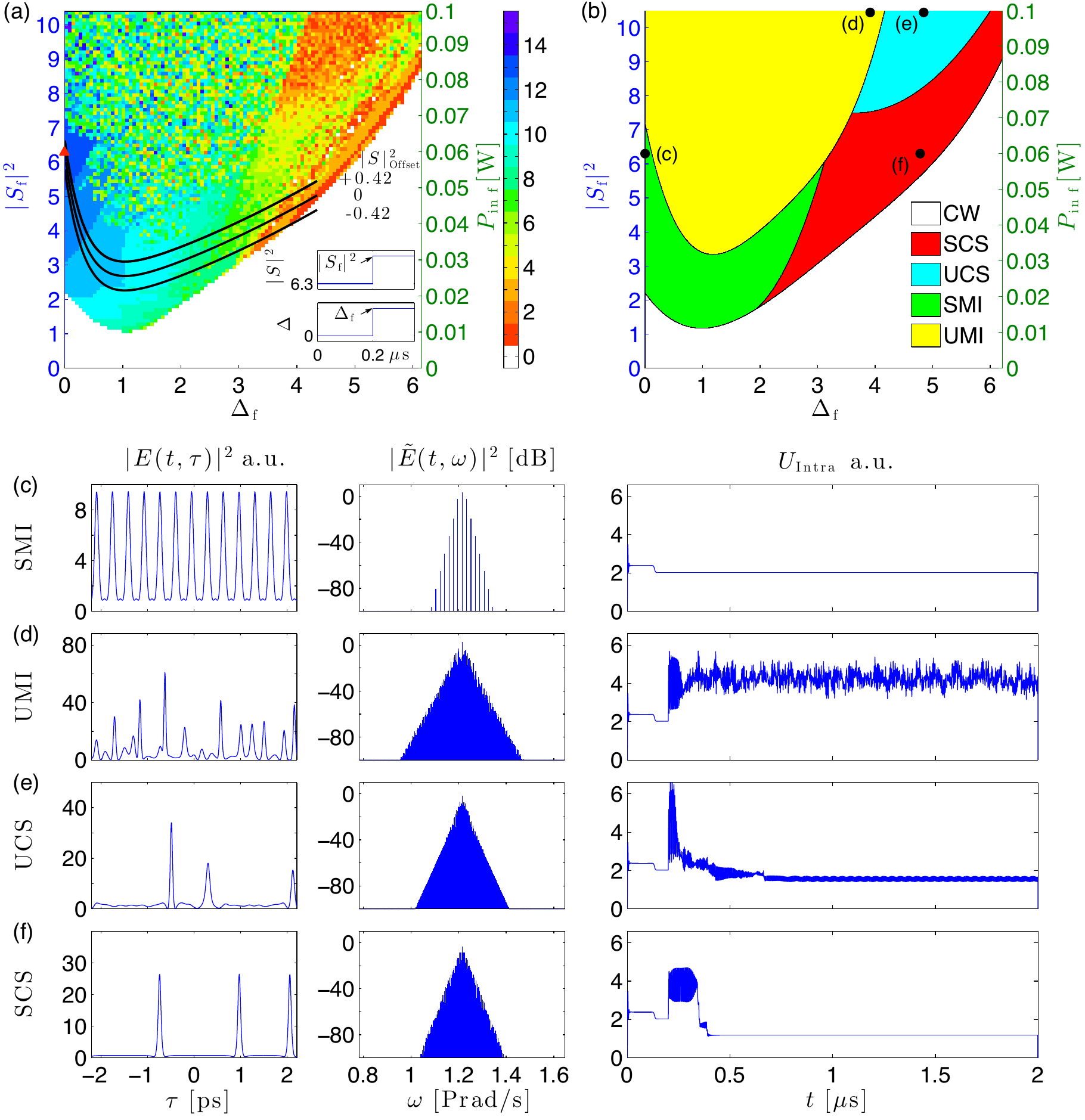}}
\caption{(a) The number of temporal peaks as a function of the final point in $(\Delta,|S|^2)$ parameter space at low pump power. The red triangle shows the initial point of the simulations. These simulations were done using the same realization of initial noise. The black curves correspond to CATs using Eq. 4 with $|S|^2_{\textrm{\fontsize{6}{1}\selectfont{Offset}}}$ of -0.42, 0, and 0.42 equivalent to offsets of -4, 0 and 4 mW respectively. (b) Characterized regions according to the features of each region shown in (c) to (f). (c-f) Final intensity (left), spectrum (center) and intracavity energy versus time $t$ (right) for the point (c) $(\Delta_{\textrm{\fontsize{6}{1}\selectfont{f}}}=0,|S_{\textrm{\fontsize{6}{1}\selectfont{f}}}|^2=6.3)$ in the SMI region, (d) $(\Delta_{\textrm{\fontsize{6}{1}\selectfont{f}}}=3.9,|S_{\textrm{\fontsize{6}{1}\selectfont{f}}}|^2=10.5)$ in the UMI region, (e) $(\Delta_{\textrm{\fontsize{6}{1}\selectfont{f}}}=4.8,|S_{\textrm{\fontsize{6}{1}\selectfont{f}}}|^2=10.5)$ in the UCS region, and (f) $(\Delta_{\textrm{\fontsize{6}{1}\selectfont{f}}}=4.8,|S_{\textrm{\fontsize{6}{1}\selectfont{f}}}|^2=6.3)$ in the SCS region.}
\label{fig:Fig2}
\end{figure}

Because instantaneously jumping to a single final state in detuning-power parameter space as in the simulations of Figs. 2(a) is not physically realizable, we next define a path that can be followed smoothly between our starting point $(\Delta = 0,|S|^2 = 6.3)$ and a target point chosen to fall within the single CS region $(\Delta = 4.35,|S|^2 = 5.04)$. In particular, we select a path that goes around the chaotic and unstable regions. We call such a path a chaos-avoiding trajectory (CAT). In this study we consider a CAT given by the following:

\begin{equation}
        |S|^2=4.15\exp(-3.09\Delta)+2.15\exp(0.196\Delta)+|S|^2_{\textrm{\fontsize{6}{1}\selectfont{Offset}}}
\label{eq:CAT}
\end{equation}

This CAT is defined using a curve-fitting tool and takes the form of a two-term exponential function. Additionally, we introduce a power offset parameter $|S|^2_{\textrm{\fontsize{6}{1}\selectfont{Offset}}}$ which allows us to describe a family of CATs, portrayed by the black curves in Fig. 2(a). To test the performance of the CAT we run simulations in three stages. In the first stage we dwell at the selected initial point $(\Delta = 0,|S|^2 = 6.3)$ for 1.0 $\mu$s ($\approx$728 $t_{\textrm{\fontsize{6}{1}\selectfont{ph}}}$) to produce a steady state Kerr comb in the SMI regime. Next, we sweep the detuning linearly from $\Delta = 0$ to a final value $\Delta_{\textrm{\fontsize{6}{1}\selectfont{f}}}$ over a 0.3 $\mu$s interval ($\approx$218 $t_{\textrm{\fontsize{6}{1}\selectfont{ph}}}$) while varying the input power according to the CAT in Eq. 4. We then continue the simulation at fixed detuning and input power for an additional 2 $\mu$s ($\approx$1445 $t_{\textrm{\fontsize{6}{1}\selectfont{ph}}}$). 

\begin{figure}[htpb]
\centerline{\includegraphics[width=31.5pc]{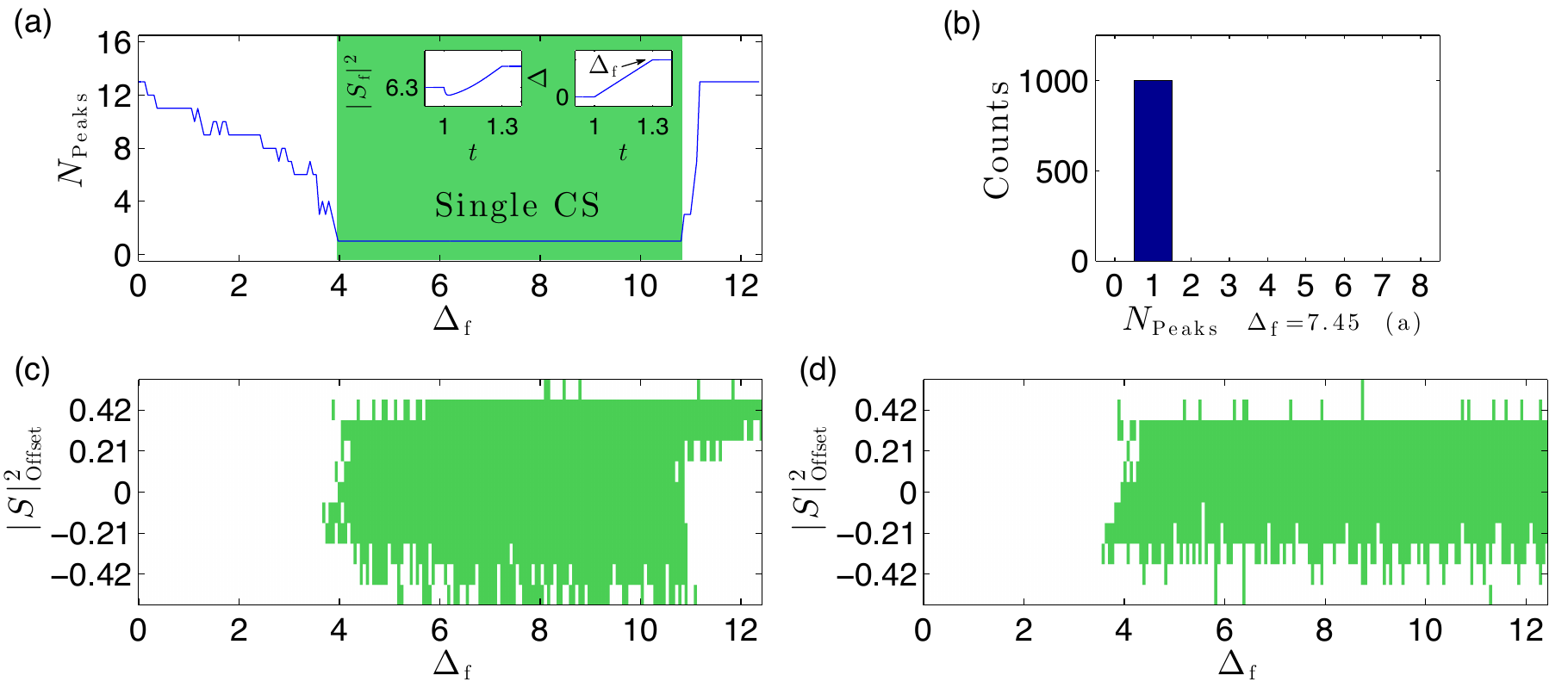}}
\caption{(a) The number of peaks as a function of final value of detuning $\Delta_{\textrm{\fontsize{6}{1}\selectfont{f}}}$ with $\Delta$ swept as shown in the inset with pump power adjusted through the CAT, Eq. 4, with $|S|^2_{\textrm{\fontsize{6}{1}\selectfont{Offset}}}= 0$  using the same realization of initial noise for all simulations. (b) Histogram of number of peaks for 1000 simulations with different realizations of initial noise with the same pump power and detuning parameters of the simulation of (a) with $\Delta_{\textrm{\fontsize{6}{1}\selectfont{f}}} = 7.45$. (c-d) Region in which a single CS is generated when a uniform offset $|S|^2_{\textrm{\fontsize{6}{1}\selectfont{Offset}}}$ is applied to the CAT for different values of final detuning $\Delta_{\textrm{\fontsize{6}{1}\selectfont{f}}}$ with (c) constant detuning interval of 0.3 $\mu$s and (d) constant detuning speed of 25 units of $\Delta$ per $\mu$s.}
\label{fig:Fig3}
\end{figure}

Figure 3(a) shows the number of temporal peaks versus the final value $\Delta_{\textrm{\fontsize{6}{1}\selectfont{f}}}$ for fixed realization of initial noise. The insets illustrate the coordinated variation of $\Delta$ and $|S|^2$ in time according to the CAT. The green area depicts a range of $\Delta_{\textrm{\fontsize{6}{1}\selectfont{f}}}$ values (from 4 to 10.8) for which a single CS is always obtained. In another case we keep $\Delta_{\textrm{\fontsize{6}{1}\selectfont{f}}}$ fixed but repeat the simulation for 1000 different realizations of initial noise. As shown in the histogram of Fig. 3(b), a single CS is obtained every time. Fig. 3(c) is similar to Fig. 3(a), in that we keep the noise initialization fixed and vary $\Delta_{\textrm{\fontsize{6}{1}\selectfont{f}}}$, except that now we use different values of $|S|^2_{\textrm{\fontsize{6}{1}\selectfont{Offset}}}$ from -0.52 to 0.52 equivalent to an offset of -5 to 5 mW to the CAT. The green shaded area again shows the cases where we generate a single CS. Additionally, we test the CAT performance keeping the sweep rate of the detuning constant at 25 units of $\Delta$ per $\mu$s for different values of $\Delta_{\textrm{\fontsize{6}{1}\selectfont{f}}}$.  This is in contrast to the simulation of Fig. 3(c) in which the detuning in swept over a constant interval of 0.3 $\mu$s, implying different sweep rates for different final detunings. The simulation results are displayed in Fig. 3(d), which again shows a green area in which a single CS state is reached. Moreover, we perform simulations using half and twice the sweep rate of the detuning with very similar results of Fig. 3(d). All these simulations demonstrate that the CAT for repeatable generation of a single CS is nonunique and can be robust over a finite range of operation. It should be emphasized that this CAT could take other forms different than a two-term exponential function, provided that it avoids the unstable and chaotic regions.

We note that thermal nonlinearities play an important role in practical  microresonators. The response time of such thermal nonlinearities is much slower than that of the Kerr nonlinearity, which complicates computational studies. For this reason thermal nonlinearity is not included in the simulations in this paper. Further research is required to determine whether the current scheme for deterministic generation of single cavity solitons remains effective with the inclusion of thermal nonlinearity. At a minimum we anticipate that it may be necessary to slow down the CAT sweep speed to a time scale consistent with the thermal nonlinearity.

\section{CS compression}

\begin{figure}[htpb]
\centerline{\includegraphics[width=31.5pc]{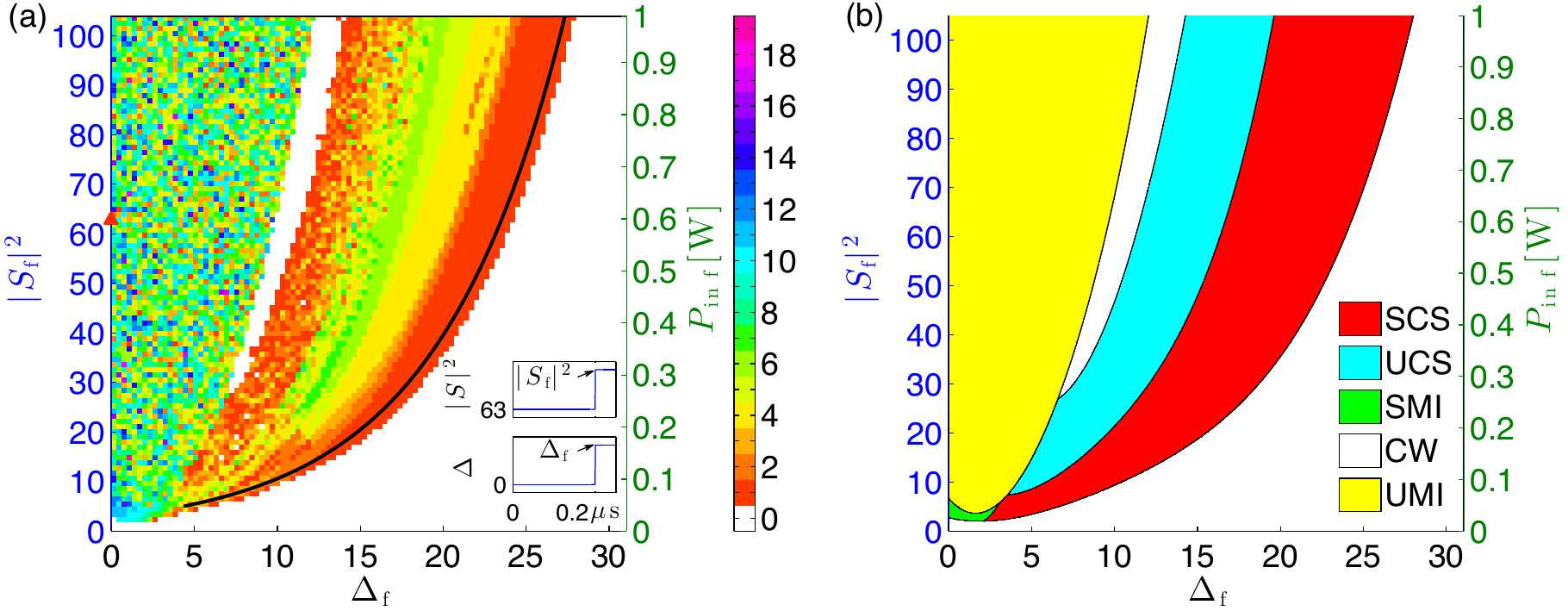}}
\caption{The number of temporal peaks as a function of the final point in $(\Delta,|S|^2)$ parameter space at high pump power. The red triangle shows the initial point of the simulations. These simulations were done using the same realization of initial noise. The black curve corresponds to a compression function using Eq. 5. The initial point of this compression function is the final point of the CAT with $|S|^2_{\textrm{\fontsize{6}{1}\selectfont{Offset}}}= 0$ in Fig. 2(a). (b) Characterized regions using the same criteria of Fig. 2(b).}
\label{fig:Fig4}
\end{figure}

Our final goal is to demonstrate compression behavior when we move the system from relatively low to relatively high detuning and power under a trajectory that remains within the SCS region. A process similar to the characterization of Fig. 2 is performed to explore behavior over a larger range of parameter space (final values of detuning $\Delta$ and input power $|S|^2$ up to 31 and 105 (1 W), respectively), starting at initial point of detuning $\Delta = 0$ and pump power $|S|^2= 63$  (0.6 W). This point has higher power than the initial point used in Fig. 2(a) because in this higher range of power, the microresonator needs more initial energy to show all the possible regions of behavior without collapsing to the CW state. The results of the simulations in this range and the classification are shown in Figs. 4(a) and 4(b), respectively. Then, we define a compression function from $(\Delta = 4.35,|S|^2 = 5.04)$, the end of a previous CAT trajectory, to $(\Delta = 27.45,|S|^2 = 105)$ using a trajectory given by
\begin{equation}
        |S|^2=2.846\exp(0.1316\Delta)
\label{eq:CAT}
\end{equation}
Again, this function is not unique, and a wide range of end points $\Delta$ and $|S|^2$ lead to similar results, provided that the trajectory remains within the SCS region. 

To provide a complete picture of the generation and compression processes of a single CS, we run a simulation in five stages, as shown in Fig. 5(a), which plots the coordinated change in detuning and input power versus slow time. Figure 5(e) shows the intensity and spectrum of the steady state initial point (Turing rolls pattern of 13 temporal peaks). In Fig. 5(b) the Turing rolls pattern fall into a single CS around 1.4 $\mu$s. Fig 5(f) shows the intensity and spectrum of the generated single CS. In Fig. 5(c) we can observe how the Kerr comb is broadening while the system is going through the compression function between 1.6 $\mu$s and 2.6 $\mu$s. Figures 5(f) and (g) show that the CS is compressed from 73 fs to 29 fs full width at half maximum (FWHM) pulse width and the bandwidth $\Delta\omega/2\pi$ goes from 3.98 to 10.2 THz. Figure 5(d) shows $U_{\textrm{\fontsize{6}{1}\selectfont{Intra}}}$ along a CAT leading to single CS generation and subsequent compression. Neither noisy nor oscillatory behavior is seen, providing evidence that the system remains in SMI or SCS regions for all simulation time and should consequently be highly stable with low intensity noise.

Although we are not taking into account high order dispersion terms and other nonlinear effects (e.g. Raman) that will be present in real systems, we believe these results make clear that once a single CS state is reached compression should be possible while maintaining stability through appropriate coordinated increases in $\Delta$ and $|S|^2$.

\begin{figure}[htpb]
\centerline{\includegraphics[width=31.5pc]{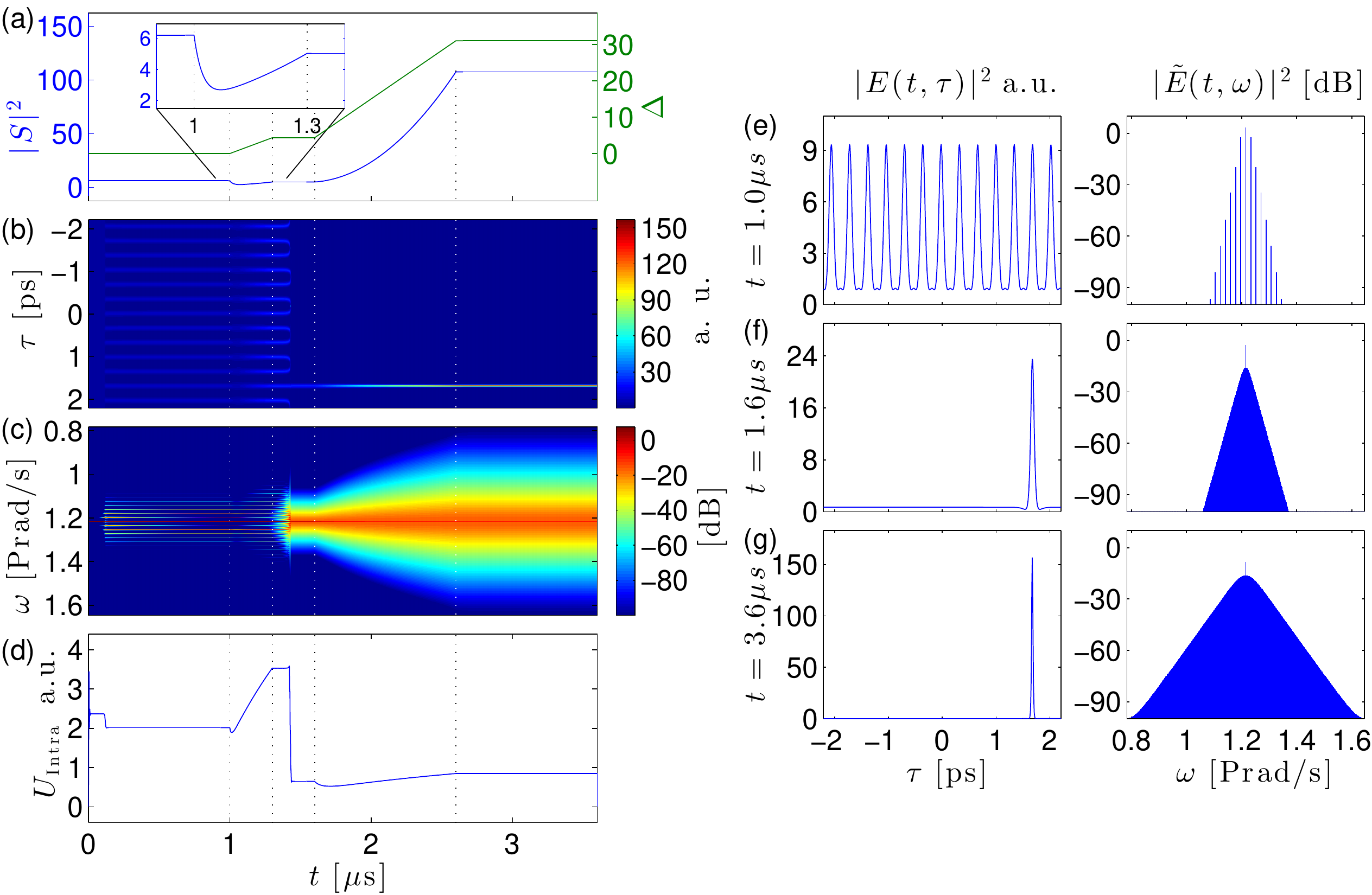}}
\caption{Generation and compression processes of a single CS through the CAT and compression functions in Eq. 4 and Eq. 5, respectively. From top to bottom: (a) pump power (blue) and detuning (green), (b) temporal intensity, (c) spectrum, (d) total intracavity energy vs. slow time $t$. (e-g) Intensity (right) and spectrum (left) at (e) $t = 1$ $\mu$s, steady state initial point, (f) $t = 1.6$ $\mu$s, after single CS generation and (g) $t = 3.6$ $\mu$s, after compression.}
\label{fig:Fig5}
\end{figure}

\section{Conclusions}

In this work, we have presented a novel method to generate a single CS in anomalous dispersion microresonators in a highly deterministic way through coordinated tuning of pump frequency and power in order to avoid chaotic and unstable operating regimes. The simulation results presented here could help accelerate progress towards establishing microring resonators as highly coherent and stable single FSR Kerr comb sources for practical applications. Furthermore, we have also demonstrated a way to compress the single CS by further coordinated variation of the pump parameters in the region where CSs are stable.

\section*{Acknowledgments}

This work was supported in part by the National Science Foundation under grant ECCS- 1102110, by the Air Force Office of Scientific Research under grant FA9550-12-1-0236, and by the DARPA PULSE program through grant W31P40- 13-1-0018 from AMRDEC. JAJ acknowledges support by Colciencias Colombia through the Francisco Jose de Caldas Conv. 529 scholarship and Fulbright Colombia. JAJ is grateful to Victor Torres-Company and Evgenii Narimanov for fruitful discussions and to Joseph Lukens for a careful reading of this manuscript.

\end{document}